\documentclass[9pt,conference]{IEEEtran}

\usepackage{waspaa25}

\usepackage{bm} 
\usepackage{tabularray}
\usepackage{multirow}
\usepackage{musicography}
\usepackage{booktabs}
\usepackage{graphicx} 
\usepackage{xcolor}

\definecolor{forestgreen}{rgb}{0.18,0.55,0.34}

\title{Multi-Class-Token Transformer for\\ Multitask Self-supervised Music Information Retrieval}

\name{Yuexuan Kong$^{1, 2}$,
      Vincent Lostanlen$^{2}$,      
      Romain Hennequin$^{1}$,
      Mathieu Lagrange$^{2}$,
      Gabriel Meseguer-Brocal$^{1}$}
\address{$^{1}$Deezer Research, Paris, France \;
$^{2}$Nantes Université, Centrale Nantes, CNRS, LS2N, UMR 6004,
F-44000 Nantes, France
}

\begin{document}
\maketitle

\begin{abstract}
Contrastive learning and equivariant learning are effective methods for self-supervised learning (SSL) for audio content analysis.
Yet, their application to music information retrieval (MIR) faces a dilemma: the former is more effective on tagging (e.g., instrument recognition) but less effective on structured prediction (e.g., tonality estimation); The latter can match supervised methods on the specific task it is designed for, but it does not generalize well to other tasks.
In this article, we adopt a best-of-both-worlds approach by training a deep neural network on both kinds of pretext tasks at once.
The proposed new architecture is a Vision Transformer with 1-D spectrogram patches (ViT-1D), equipped with two class tokens, which are specialized to different self-supervised pretext tasks but optimized through the same model: hence the qualification of self-supervised multi-class-token multitask ($\text{MT}^2$).
The former class token optimizes cross-power spectral density (CPSD) for equivariant learning over the circle of fifths, while the latter optimizes normalized temperature-scaled cross-entropy (NT-Xent) for contrastive learning.
$\text{MT}^2$ combines the strengths of both pretext tasks and outperforms consistently both single-class-token ViT-1D models trained with either contrastive or equivariant learning. 
Averaging the two class tokens further improves performance on several tasks, highlighting the complementary nature of the representations learned by each class token.
Furthermore, using the same single-linear-layer probing method on the features of last layer, $\text{MT}^2$ outperforms MERT on all tasks except for beat tracking; achieving this with 18x fewer parameters thanks to its multitasking capabilities.
Our SSL benchmark demonstrates the versatility of our multi-class-token multitask learning approach for MIR applications.
\end{abstract}

\section{Introduction}
\label{sec:intro}
Self-supervised learning (SSL) consists in designing a \emph{pretext task} to train deep neural networks from unlabeled data.
The promise of SSL is that, although this task does not reflect a real-world use case, its resolution is transferable to \emph{downstream tasks} with little or no supervised fine-tuning \cite{balestriero2023cookbook}.
SSL has recently found many applications in the context of music information retrieval (MIR) \cite{meseguer2024experimental}.

Contrastive learning is arguably the simplest kind of SSL, with CLMR \cite{spijkervet2021contrastive} and MULE \cite{mccallum2022supervised} being two examples for MIR.
Given an audio segment, named \emph{anchor}, these methods extract a \emph{positive sample}, i.e., a neighboring segment from the same music track or an artificially modified version of the anchor.
Segments from a distinct music track from the anchor are regarded as \emph{negative samples}.
Recent improvements to contrastive learning have focused on improving the random sampling of negatives in terms of effectiveness and interpretability \cite{guinotloev,choi2022towards}.
Yet, because it assigns the same importance to all negatives, contrastive learning is inefficient for structured downstream tasks such as tonality estimation \cite{downie2010music}.

Equivariant learning consists in regressing the parameter underlying a certain artificial transformation by learning to correlate two transformed copies of the same unlabeled audio segment.
Crucially, these artificial transformations are tailored to the downstream task: e.g., time shifts for beat tracking \cite{desblancs2023zero}, pitch shifts for fundamental frequency estimation \cite{gfeller2020spice,riou2023pesto}, time warps for tempo estimation \cite{quinton2022equivariant}, and so forth.
Compared to contrastive learning, equivariant learning eliminates the need to carefully craft positive and negative samples for each anchor.
Equivariant pretext tasks have the drawback of reducing the representation to a single degree of freedom.

Masked language modeling (MLM) involves predicting a masked word based on its context. Originally introduced as BERT in NLP, this pretext task has been adapted to music in models like MERT \cite{li2023mert} and MusicFM \cite{won2024foundation}. While effective on downstream tasks, these models rely on complex SSL techniques, such as teacher-student distillation with exponential moving averages, which require costly, trial-and-error hyperparameter tuning.
Furthermore, MLM requires large neural network architectures: e.g., 95M parameters for MERT.

Meanwhile, there is a pressing need for multitask MIR  \cite{weston2011multi,hamel2013transfer,hu2014mirex,scholz2016cross,hung2019multitask,kim2020one,singh2021multitask,turian2022hear}.
Yet, SSL for MIR faces a conundrum: equivariant learning achieves state-of-the-art (SOTA) performance on structured downstream tasks, contrastive learning excels in others, and MLM models reach SOTA performance across most tasks, though at a high computational cost.

In this article, we resolve this conundrum by training a neural network with only 5.3M parameters on multiple pretext tasks at once: i.e., a contrastive and an equivariant learning task.
The key idea is to compute loss functions of each self-supervised pretext task over different \emph{class tokens}, which are prepended to the \emph{sequence tokens} of a Vision Transformer with 1-D patches (ViT-1D) \cite{kong2025emergent}.
Hence the proposed name: multi-class-token multitask ($\text{MT}^2$) ViT-1D.
By averaging both class tokens for downstream tasks, $\text{MT}^2$ outperforms single-class-token monotask ViT-1D models with the same number of parameters, each optimized with either contrastive or equivariant loss, and the MLM model MERT's last-layer representation while having 18x fewer parameters.
Interestingly, the advantage of $\text{MT}^2$ is not limited to class tokens but is also observable in sequence tokens, which outperform MERT in chord estimation. The code and model weights are publicly available at \href{https://github.com/deezer/mt2}{https://github.com/deezer/mt2}. 

\section{Methods}
\begin{figure*}
\centering
\includegraphics[width=0.9\textwidth]{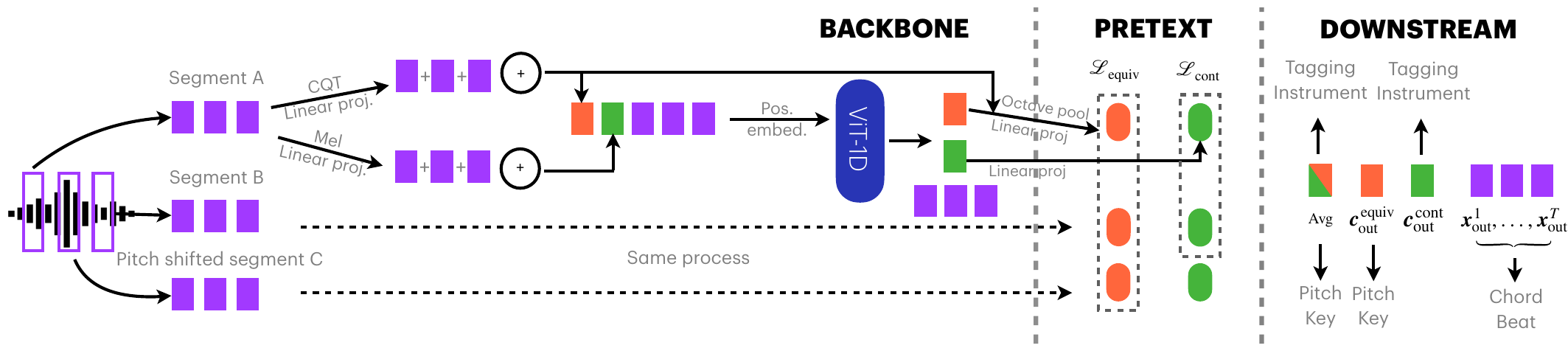}
\caption{Overview of the proposed method for mult-class-token multitask self-supervised learning with Vision Transformer with 1-D patches ($\text{MT}^2$).}
\label{fig:flowchart}
\end{figure*}

Our key contributions, multi-class-token multitasking ($\text{MT}^2$) and token-based downstream usage, are detailed in Sections \ref{sub:mt2} and \ref{sub:averaging}, with an overview in Figure \ref{fig:flowchart}. ViT-Mel (Section \ref{sub:vit-mel}) and ViT-CQT (Section \ref{sub:vit-cqt}), components of $\text{MT}^2$, also serve as baselines in Section \ref{sec:results}.

\label{sec:methods}

\subsection{Vision Transformer with mel-frequency patches (ViT-Mel)}
\label{sub:vit-mel}
ViT-Mel processes audio segments of four seconds, sampled at $16$ kHz. We compute a mel-frequency spectrogram with $128$ bins and a frame rate of \SI{31.5}{\hertz}. This representation is commonly used in MIR for contrastive learning in tagging tasks \cite{meseguer2024experimental, mccallum2022supervised}.
Each spectrogram frame is a 1-D vector, indexed by mel-frequency. We normalize this vector and apply a learnable linear layer, producing a patch sequence $\boldsymbol{x}_\mathrm{mel} = [\boldsymbol{x}_{1}, ..., \boldsymbol{x}_{T}]$ in dimension $d_{\mathrm{e}} = 192$, with $T=126$ being the sequence length.

We build a Vision Transformer with 1-D patches (ViT-1D) \cite{kong2025emergent} with an embedding dimension of 192, a depth of 12 blocks, and three attention heads per block, which is the smallest ViT version proposed in the original paper
\cite{dosovitskiy2021an}.
We prepend a learnable \emph{class token} $\boldsymbol{c}^{\mathrm{in}}_\mathrm{cont}$ to the patch sequence \cite{caron2021emerging}, defined as:
\vspace{-1.5ex}
\begin{equation}
\boldsymbol{c}_\mathrm{cont}^{\mathrm{in}} = \boldsymbol{\tilde{c}}_\mathrm{cont}^{\mathrm{in}} + \dfrac{1}{T} \sum_{t=1}^{T} \boldsymbol{x}^{mel}_{t}.
\end{equation}
where $\boldsymbol{\tilde{c}}^{\mathrm{in}}_\mathrm{cont}$ are learnable parameters, and ``cont'' refers to ``contrastive'', detailed in Section \ref{sub:nt-xent}.

This yields the input sequence as $[\boldsymbol{c}^{\mathrm{in}}_\mathrm{cont},\boldsymbol{x}^{\mathrm{in}}_{1},\ldots,\boldsymbol{x}^{\mathrm{in}}_{T}]$. Likewise, the output sequence has the form $[\boldsymbol{c}^{\mathrm{out}}_\mathrm{cont},\boldsymbol{x}^{\mathrm{out}}_{1},\ldots,\boldsymbol{x}^{\mathrm{out}}_{T}]$.
We apply a 2-D positional encoding over time and mel-frequency to the input sequence.
In the ViT-1D, the \emph{class token} is processed in the same way as sequence tokens, with shared weight in the MLP layers following the attention blocks.
The \emph{class token} serves as a summarization of the sequence, as proposed in \cite{dosovitskiy2021an}.
Before applying the loss, a learnable linear layer maps $\boldsymbol{c}^{\mathrm{out}}_\mathrm{cont}$ onto a vector $\boldsymbol{z}_\mathrm{cont}$ of dimension 512.

\subsection{Normalized temperature-scaled cross-entropy (NT-Xent)}
\label{sub:nt-xent}
We train ViT-Mel with self-supervised contrastive learning.
We extract two disjoint segments $A$ and $B$ from the same song and regard them as an anchor and a positive sample.
In the same batch, extracted segments from other songs are considered as negative samples for this anchor\cite{mccallum2022supervised}.
The contrastive loss function is normalized temperature-scaled cross-entropy (NT-Xent), defined as
:
\vspace{-1.5ex}
\begin{equation}
    \mathcal{L}_{A, B}^{\mathrm{cont}}(\boldsymbol{w}) = - \log \frac{\exp(\text{sim}(\boldsymbol{z}_{A}^\mathrm{cont}, \boldsymbol{z}_{B}^\mathrm{cont}) / \tau)}{
    \sum_{k\neq B} \exp(\text{sim}(\boldsymbol{z}_{A}^\mathrm{cont}, \boldsymbol{z}_{k}^\mathrm{cont}) / \tau)}
\end{equation}
where $\mathrm{sim}$ denotes the cosine similarity, $\tau=0.1$ is a temperature hyperparameter, and $\boldsymbol{w}$ are the trainable weights of ViT-Mel.

\subsection{Vision Transformer with CQT patches (ViT-CQT)}
\label{sub:vit-cqt}
Alternatively, we build a ViT on the constant-$Q$ transform (CQT) with $Q=12$ bins per octave over $J=8$ octaves. CQT is commonly used as representations for harmonic related tasks \cite{kong2024stone, kongskey, Bittner:DeepSalience:ISMIR:17}.
We make the same design choices as for ViT-Mel regarding embedding dimension $d_{\mathrm{e}}$, sequence length $T$, ViT-1D architecture (Section \ref{sub:vit-mel}). The CQT input sequence is denoted as $\boldsymbol{x}_{\mathrm{cqt}} = [\boldsymbol{x}_{1}, ..., \boldsymbol{x}_{T}]$.
We prepend a \emph{class token} in the same way as for ViT-Mel, initialized with learnable parameters and the average of $\boldsymbol{x}_\mathrm{cqt}$.

At the output of ViT-1D, a residual connection is added from the average of $\boldsymbol{x}_\mathrm{cqt}$ to $\boldsymbol{c}^\mathrm{out}_\mathrm{equiv}$(``equiv'' refers to ``equivariant'', detailed in Section \ref{sub:cpsd}), obtaining:
\vspace{-1em}
\begin{equation}
\boldsymbol{\tilde{c}}^{\mathrm{out}}_\mathrm{equiv} = \boldsymbol{c}^{\mathrm{out}}_\mathrm{equiv} + \dfrac{1}{T} \sum_{t=1}^{T} \boldsymbol{x}_{t}^\mathrm{cqt}.
\end{equation}

We attach a learnable linear layer to project $\boldsymbol{\tilde{c}}^{\mathrm{out}}_\mathrm{equiv}$ into dimension of $QJ=84$, then we reduce the dimension to $Q=12$ via octave pooling.
This is similar to the extraction of chroma features in MIR, except that our averaging procedure operates on a learned representation as opposed to directly on the CQT.
After averaging, we apply a softmax layer, yielding a nonnegative vector $\boldsymbol{z}_\mathrm{equiv}$ in dimension 12 whose entries sum to one.

\subsection{Cross-power spectral density (CPSD)}
\label{sub:cpsd}

We train ViT-CQT with self-supervised equivariant learning. 
Cross-power spectral density (CPSD), introduced in STONE, is used to structure the embedding space in a circularly equivariant manner, resembling the circle of fifths \cite{kong2024stone}.
Let $\boldsymbol{\hat{z}}$ be the discrete Fourier transform over $\mathbb{Z}_{12}$ of $\boldsymbol{z}_\mathrm{equiv}$.
The CPSD of $\boldsymbol{z}_{A}^\mathrm{equiv}$ and $\boldsymbol{z}_B^\mathrm{equiv}$ is given by $\boldsymbol{\hat{z}}_{A}\boldsymbol{\hat{z}}_{B}^{*}$, where the asterisk denotes the complex conjugate.
Given a hyperparameter $\omega\in\mathbb{Z}$ and a pitch class interval parameter $k\in\mathbb{Z}$, we consider the following function:
\vspace{-1.5ex}
\begin{equation} 
\mathcal{D}_{k}(\boldsymbol{z}_{A}, \boldsymbol{z}_{B}) = \dfrac{1}{2} \left \vert e^{- 2\pi\mathrm{i}\omega k/12} - \widehat{\boldsymbol{z}}_{A}[\omega]\widehat{\boldsymbol{z}}_{B}^{\ast}[\omega] \right \vert^2. 
\label{eq:dft} 
\end{equation}
By setting $\omega=7$, $\mathcal{D}_{k}$ is a differentiable distance function over the circle of fifths.
Indeed, by property of the softmax and since 7 is coprime with 12, $\mathcal{D}_{k}$ is zero if and only if $\boldsymbol{z}_{A}^\mathrm{equiv}$ and $\boldsymbol{z}_{B}^\mathrm{equiv}$ have a single nonzero entry and differ by a circular shift of $k$ semitones.
Assuming that a musical piece does not modulate between segments A and B, the minimization of $\mathcal{D}_{0}(\boldsymbol{z}_{\mathrm{A}},\boldsymbol{z}_{\mathrm{B}})$ with respect to trainable weights $\boldsymbol{w}$ of ViT-CQT is an equivariant pretext task for self-supervised tonality estimation.
Moreover, assuming that a third segment $C$ is available and is known to be $k$ semitones apart from $A$ and $B$, then it is judicious to minimize $\mathcal{D}_{k}(\boldsymbol{z}_{\mathrm{A}},\boldsymbol{z}_{\mathrm{C}})$ and $\mathcal{D}_{k}(\boldsymbol{z}_{\mathrm{B}},\boldsymbol{z}_{\mathrm{C}})$.
In practice, we obtain $C$ by artificial frequency transposition of $A$ and randomize the parameter $k$ uniformly between -5 and +6 semitones.
Thus, our loss function for self-supervised equivariant learning is:
\vspace{-1.5ex}
\begin{equation}
\mathcal{L}^{\mathrm{equiv}}_{A, B, C}(\boldsymbol{w})=
\mathcal{D}_{0}(\boldsymbol{z}_{A}, \boldsymbol{z}_{B}) \nonumber
+ \mathcal{D}_{k}(\boldsymbol{z}_{A}, \boldsymbol{z}_{C}) \nonumber
+ \mathcal{D}_{k}(\boldsymbol{z}_{B}, \boldsymbol{z}_{C}).
\label{eq:cpsd}
\end{equation}
The loss function above has allowed a self-supervised convolutional network to match the supervised SOTA in the automatic classification of major and minor keys \cite{kongskey}.
To our knowledge, our article is the first to apply CPSD-based equivariant learning to Transformers, which, unlike convolutional networks, are not inherently equivariant to transpositions. The skip connection before the final linear layer is crucial for faster convergence when learning the equivariant property.

\subsection{Multi-class-token multitasking ($\text{MT}^2$)}
\label{sub:mt2}

We add patch sequences $\boldsymbol{x}_\mathrm{cqt}$ (Section \ref{sub:vit-cqt}) and $\boldsymbol{x}_\mathrm{mel}$ (Section \ref{sub:vit-mel}) to produce a patch sequence $\boldsymbol{x}_{\mathrm{p}} = \boldsymbol{x}_{\mathrm{cqt}} + \boldsymbol{x}_{\mathrm{mel}}$, to which we prepend two learnable class tokens: 
an equivariant class token $\boldsymbol{c}_{\mathrm{equiv}}^\mathrm{in} \in \mathbb{R}^{d_e}$ initialized with learnable parameters and the average of $\boldsymbol{x}_\mathrm{cqt}$, and a contrastive class token $\boldsymbol{c}_{\mathrm{cont}}^\mathrm{in} \in \mathbb{R}^{d_e}$ initialized with learnable parameters and the average of $\boldsymbol{x}_\mathrm{mel}$.
Thus, the input token sequence to the transformer consists of the concatenation of the two class tokens followed by the frame-wise patch sequence $\boldsymbol{x}_{\mathrm{p}}$. We use the same ViT-1D architecture as ViT-Mel and ViT-CQT: therefore, no additional parameters are introduced when training $\text{MT}^2$. Both class tokens are updated through the same attention layers and MLP layers, with no constraints beyond their respective losses, allowing each to selectively aggregate information.

The output token sequence of the last layer is denoted as $[\boldsymbol{c}_{\mathrm{out}}^{\mathrm{equiv}}, \boldsymbol{c}_{\mathrm{out}}^{\mathrm{cont}}, \boldsymbol{x}_{\mathrm{out}}^{1}, ..., \boldsymbol{x}_{\mathrm{out}}^{T}]$ where $\boldsymbol{c}_{\mathrm{out}}^{\mathrm{equiv}}$ and $\boldsymbol{c}_{\mathrm{out}}^{\mathrm{cont}}$ correspond to the output equivariant and contrastive tokens respectively. 
Before applying the losses, we have two distinct heads for the two class tokens, same as described in Section \ref{sub:vit-mel} and \ref{sub:vit-cqt}. $\boldsymbol{c}_{\mathrm{out}}^{\mathrm{equiv}}$ added with its residual connection from $\boldsymbol{x}_\mathrm{cqt}$ is passed through an octave pooling layer and softmax layer obtaining an output of 12 dimensions $\boldsymbol{z}^{\mathrm{equiv}}$, while $\boldsymbol{c}_{\mathrm{out}}^{\mathrm{cont}}$ is linearly projected into a space of 512 dimensions $\boldsymbol{z}^{\mathrm{cont}}$.
 
We optimize $\boldsymbol{z}^\mathrm{equiv}$ using an equivariant loss $\mathcal{L}^{\mathrm{equiv}}$, and $\boldsymbol{z}^{\mathrm{cont}}$ using a contrastive loss $\mathcal{L}^{\mathrm{cont}}$ applied across the batch.
Recent publications in computer vision have proposed similar transformer architectures in which multiple learnable tokens are prepended to the beginning of the sequence.
\cite{darcet2024vision} introduced register tokens, which are discarded during downstream training but serve to enhance the emergent properties observed in attention maps.
MCTransformer+ \cite{xumct} uses multiple class tokens with distinct supervision signals corresponding to different classes.
These approaches have been shown to improve both the emergent properties of transformer representations and performance on downstream tasks. 
However, to the best of our knowledge, similar approaches have not yet been explored in a fully self-supervised manner, with different SSL losses, particularly in MIR.
Unlike above work, both class tokens in $\text{MT}^2$ are optimized in a fully self-supervised manner, structured differently only through their respective losses, and both are required for downstream tasks.

\subsection{Usage of tokens for downstream tasks}
\label{sub:averaging}
The output sequence of $\text{MT}^2$ consists of two class tokens, $\boldsymbol{c}_{\mathrm{out}}^{\mathrm{equiv}}$ and $\boldsymbol{c}_{\mathrm{out}}^{\mathrm{cont}}$, along with sequence tokens, $[\boldsymbol{x}_{\mathrm{out}}^{1}, ..., \boldsymbol{x}_{\mathrm{out}}^{T}]$.
For downstream task training, we average both class tokens, as they are not explicitly trained to be disentangled and may capture complementary information. 
For comparison, we also train on specific tokens for corresponding tasks, as described in Section \ref{sec:results}.
Moreover, the sequence tokens exhibit emergent properties, as noted in computer vision\cite{caron2021emerging} and music information retrieval\cite{kong2025emergent}.
Although these tokens are not directly optimized by the loss functions, we hypothesize that they capture frame-level information that both class tokens rely on to generate representations that can be optimized by losses, thus useful for downstream tasks.

\section{Application to music information retrieval}
\label{sec:results}
\subsection{Training details}
We curate a subset of 100k songs from the catalog of a commercial music streaming service. We use a batch size of 128 pairs of 4-second segments, a base learning rate of $1 \times 10^{-4}$ with a cosine decay until $5 \times 10^{-7}$, and train for 600 epochs with 512 steps per epoch.
For all downstream tasks, we train only a single linear layer on top of the 192-dimensional class or sequence tokens, keeping the ViT-1D backbone frozen.

\subsection{Global tasks}
\label{sub:global-tasks}
Global tasks are time-invariant: they require a single prediction per audio excerpt.
We consider four well-known global tasks in MIR: multilabel music tagging, instrument recognition on isolated notes, key estimation, and pitch estimation on isolated notes.

\textbf{Music tagging.} We train, validate, and test on MagnaTagATune \cite{law2009magnatagatune} with the same split as \cite{lee2017sample}. MagnaTagATune contains 25k songs from 230 artists, with multilabel annotation from 50 English tags. We probe on the contrastive class token $\boldsymbol{c}_{\mathrm{out}}^{\mathrm{cont}}$ and the average of both class tokens for $\text{MT}^2$, and the class token of both single-token monotask models (ViT-Mel and ViT-CQT). We evaluate the area under the receiver operating characteristic curve (ROC-AUC) and mean average precision (mAP) in their macro-aggregated versions.

\textbf{Instrument recognition.} We train, validate, and test on TinySOL \cite{cella2020orchideasol} with a random 8:1:1 split. TinySOL contains 3k isolated notes from 14 instruments. We use the same probing method as for music tagging. We evaluate top-1 accuracy over 14 classes.

\textbf{Key estimation.} We train and validate on FMAKv2 \cite{kong2024stone} with a random 9:1 split.
FMAKv2, an improved version of FMAK \cite{wong2023fmak}, contains 5k songs from the Free Music Archive \cite{defferrard2017fma}.
We test on GiantSteps\cite{knees2022giansteps}, which contains 604 electronic dance music tracks. We probe on the equivariant class token $\boldsymbol{c}_{\mathrm{out}}^{\mathrm{equiv}}$ and the average of both tokens for $\text{MT}^2$, and the class token of the both single-token monotask models. We evaluate the MIREX score, a weighted accuracy which assigns a lower penalty to predictions which are harmonically related to the ground truth  \cite{raffel2014mir_eval}.

\textbf{Pitch estimation.} We use TinySOL as for instrument recognition \cite{cella2020orchideasol}. We use the same probing method as in key estimation. We evaluate top-1 classification accuracy over 82 semitone intervals.

Table \ref{tab:global-tasks} summarizes our results for all four global tasks.

\begin{table}[t]
    \centering
    \footnotesize
    \setlength{\tabcolsep}{4pt} 
    \caption{Benchmark of four global tasks: multilabel music tagging, instrument recognition on isolated notes, key estimation, and pitch estimation on isolated notes. The top two rows are results of probing the class token of monotask single-token models: ViT-Mel (Section \ref{sub:vit-mel}) with contrastive learning (Section \ref{sub:nt-xent}) and ViT-CQT (Section \ref{sub:vit-cqt}) with equivariant learning (Section \ref{sub:cpsd}). The next three rows are our multi-token multitask model ($\text{MT}^2$, Section \ref{sub:mt2}), after probing the contrastive token, the equivariant token, or an averaging of the two tokens (Section \ref{sub:averaging}). The last row is probing of MERT's last layer representations, a pretrained masked language model (MLM) for music. See Section \ref{sub:global-tasks} for details about tasks and metrics.}
    \label{tab:global-tasks}
    \begin{tabular}{@{}ll|ccccc@{}}
        \toprule
        \textsc{\textbf{Model}} & \textsc{\textbf{Token}} &
        \multicolumn{2}{c}{\textsc{\textbf{Tagging}}} & \textsc{\textbf{Instrument}} & \textsc{\textbf{Key}} & \textsc{\textbf{Pitch}} \\
        & & \textsc{mAP} & \textsc{ROC} & \textsc{Acc} & \textsc{MIREX} & \textsc{Acc} \\
        \midrule
        ViT-Mel   & Cont.   & 0.334 & 0.852 & 0.904 & 0.577 & 0.973 \\
        ViT-CQT  & Equiv.  & 0.229 & 0.779 & 0.548 & 0.733 & 0.973 \\
        \midrule
        $\text{MT}^2$    & Cont.   & 0.390 & 0.884 & 0.925 & --    & -- \\
        $\text{MT}^2$    & Equiv.  & --    & --    & -- & 0.700 & 0.983 \\
        $\text{MT}^2$    & Avg.  & 0.388    & 0.884    & 0.918 & 0.713 & 0.990 \\
        \midrule
        \textsc{MERT}    & -     & 0.345 & 0.852 & 0.493 & 0.527 & 0.979 \\
        \bottomrule
    \end{tabular}
\end{table}

\subsection{Local tasks}
\label{sub:local-tasks}
Local tasks are time-equivariant: they require a prediction per frame with a frame rate typically higher than \SI{1}{\hertz}. We consider two well-established local tasks in MIR: beat tracking and chord estimation. Here, instead of probing on the class tokens, we probe on the sequence tokens $[\boldsymbol{x}_{\mathrm{out}}^{1}, ..., \boldsymbol{x}_{\mathrm{out}}^{T}]$ for ViT-Mel, ViT-CQT and $\text{MT}^2$. These sequence tokens are not directly optimized using any pretext task losses; however, they exhibit emergent properties that can be used for local tasks.

\textbf{Beat tracking.} We train and validate on the Ballroom dataset \cite{gouyon2004ballroom} with a 9:1 split. This dataset contains 698 songs. We test on GTZAN Rhythm \cite{marchand2015gtzanrhythm}, which contains 998 songs. Since beat tracking typically requires a higher frame rate than \SI{31.5}{\hertz}, we attach two independent heads to each $\boldsymbol{x}_\mathrm{out}$, doubling the frame rate to \SI{63}{\hertz}. Additionally, we apply a standard smoothing method for beat tracking, where we increase the values of the two neighboring frames to 0.5 instead of 0.
We evaluate the $F_1$-score with a tolerance window of $\pm$\SI{70}{\milli\second}.

\textbf{Chord estimation.} We collect 124 songs from the Real World Computing Pop (RWC-POP) and Schubert Winterreise Dataset (SWD) \cite{weiss2021schubert, goto2002rwc}, limited to one performance per song.
We train, validate, and test on this corpus with a random 8:1:1 split.
We consider 24 classes of major and minor chords; exclude those that cannot be reduced to these classes (e.g., suspended chords); and include a ``no chord'' class, resulting in 25 classes.
We evaluate frame-level classification accuracy at a rate of \SI{31.5}{\hertz}.

Table \ref{tab:local-tasks} summarizes our results for both local tasks.

\begin{table}[t]
    \centering
    \footnotesize
    \caption{Benchmark on two local tasks: beat tracking and chord estimation, based on probing sequence tokens. See Table \ref{tab:global-tasks} for description of models. See Section \ref{sub:local-tasks} for details about tasks and metrics.}
    \label{tab:local-tasks}
    \begin{tabular}{ll|cc}
        \toprule
        \textsc{\textbf{Model}} & \textsc{\textbf{Token}} & \textsc{\textbf{Beat}} & \textsc{\textbf{Chord}} \\
        & & \textsc{$F_1$} & \textsc{Acc} \\
        \midrule
        ViT-Mel   & Seq.   & 0.656 & 0.323 \\
        ViT-CQT  & Seq.   & 0.515 & 0.542 \\
        \midrule
        $\text{MT}^2$    & Seq.   & 0.698 & 0.447 \\
        \midrule
        \textsc{MERT}    & -    & 0.786 & 0.392 \\
        \bottomrule
    \end{tabular}
\end{table}

\section{Discussion}
\subsection{Comparison with single-token monotask models}
We compare $\text{MT}^2$ with two single-token monotask baseline models described in Section \ref{sec:methods}: ViT-Mel and ViT-CQT.
ViT-Mel is a ViT-1D whose class token is initialized with a Mel-frequency spectrogram and trained using contrastive loss.
ViT-CQT, by contrast, uses a CQT initialization and is optimized with the cross-power spectral density (CPSD) loss, an equivariant objective.

Compared to ViT-Mel, the equivariant class token of $\text{MT}^2$ shows large improvements on harmonically structured global tasks such as key and pitch estimation.
For tagging and instrument recognition, which are better suited to contrastive pretraining, the contrastive class token of $\text{MT}^2$ still outperforms ViT-Mel, despite both being optimized with the same loss.
This suggests that the multitask, multi-token framework enables beneficial inductive sharing between the two learning objectives, allowing the contrastive token to leverage information learned by the equivariant token.

ViT-CQT, being tailored for equivariant tasks through CPSD optimization, achieves strong results on harmonic tasks.
However, its performance on other tasks is substantially lower, reflecting a known limitation of equivariant SSL: task-specific representations often do not generalize well.
In contrast, while $\text{MT}^2$ shows a slight performance drop on key and chord estimation relative to ViT-CQT, it achieves considerably better results across all other tasks, demonstrating a favorable trade-off between specialization and generalization.

Moreover, we observe that averaging the two class tokens leads to 1) improved performance on key and pitch estimation,  2) matching performance on tagging, and 3)  a slight drop on instrument recognition while still outperforming both single-token monotask models on this task. This suggests that the two class tokens encode complementary information and are not fully disentangled, highlighting the benefit of considering  them in combination for certain downstream scenarios.

\subsection{Comparison with MERT}
MERT is a music representation model that achieves leading results across downstream tasks\cite{li2023mert}. 
We use the public 95M-parameter backbone version (\textsc{MERT-v1-95M}). 
To ensure a fair evaluation against $\text{MT}^2$, we adopt the same datasets, splits, and probing described in Section~\ref{sec:methods}.
It is worth noting that MERT’s original paper improves performance by using downstream-task-specific weight combinations of intermediate layers, alongside with dropout and a hidden layer of dimension 512.
In contrast, we probe only a fixed layer output (the final layer)\cite{spijkervet2021contrastive,mccallum2022supervised,won2024foundation}, apply no dropout, and use a single linear layer without hidden dimensions for both models. This might result in performance drop in downstream tasks for both models, however we focus on the benefits of multitasking by directly measuring the learned representation space for all downstream tasks.

Across all evaluated downstream tasks, $\text{MT}^2$ outperforms \textsc{MERT} under the same evaluation protocol, with the exception of beat tracking, while having 18× fewer parameters and a 4× smaller embedding space.
In particular, it achieves gains in key and chord estimation, and instrument recognition.
Interestingly, while MERT performs on par with ViT-Mel on non-harmonic tasks, $\text{MT}^2$ consistently outperforms on harmonic tasks, thanks to multitask learning via the equivariant class token.
On beat tracking, $\text{MT}^2$ underperforms. This may be explained by the fact that the emergent properties in the sequence tokens are not explicitly guided by the loss functions.


\subsection{Principal component analysis}
\label{sub:pca}
\begin{figure}
\label{fig:pca}
\includegraphics[width=\linewidth]{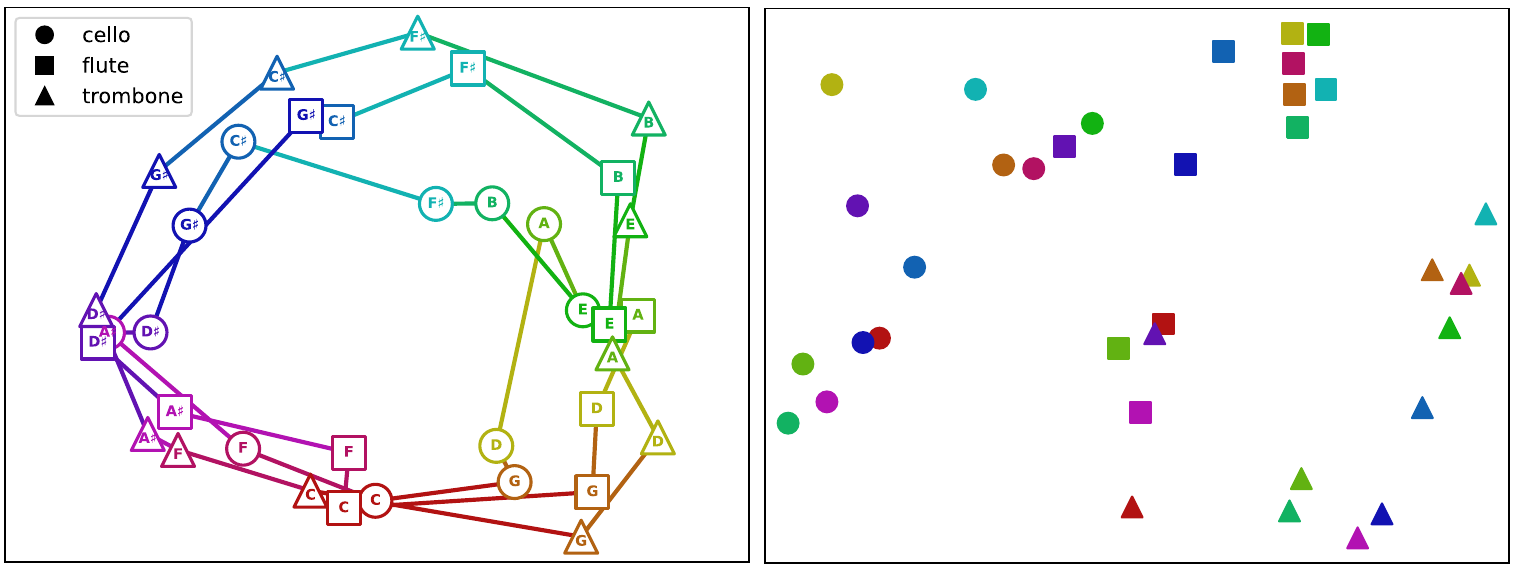}
\small
\caption {PCA of $\text{MT}^2$ embeddings using the equivariant (left) and contrastive (right) class tokens. Shapes indicate instruments; colors indicate pitches. Left: pitches cluster across instruments and form a ring aligned with the circle of fifths (CoF). Right: samples cluster by instrument, showing pitch invariance. Edges (left) connect samples from the same instrument in the order of CoF, omitted on the right figure for clarity.}
\centering
\end{figure}

To interpret the roles of each class token in $\text{MT}^2$, we select three instruments playing 12 pitches each (from $\texttt{A}_4$ to $\texttt{G}\musSharp{}_4$) from the TinySOL dataset, yielding 36 excerpts. These are processed by $\text{MT}^2$ after self-supervised training, without fine-tuning. We extract the 192-dimensional embeddings from both the contrastive and equivariant class tokens and reduce them to two dimensions using PCA. Figure~\ref{fig:pca} shows the resulting projections. 

In the left subfigure, the equivariant token embeddings form a ring structure, where similar pitches from different instruments cluster together. The angular position aligns with the circle of fifths, and points for each instrument are connected in that order. Though not a perfect dodecagon, this structure generalizes across timbral variation. The subfigure on the right represents the contrastive class token. We observe that samples cluster primarily by instrument, regardless of pitch. This indicates that the contrastive token is relatively invariant to pitch and more sensitive to timbre.

These patterns suggest that assigning distinct self-supervised losses to each class token encourages the transformer to specialize: each class token captures complementary aspects of the sequence and structures the embedding space differently.

\section{Conclusion}
In the context of self-supervised learning (SSL) for music information retrieval (MIR), we have presented $\text{MT}^2$: a simple and scalable method to train the same ViT-1D model on multiple pretext tasks at once.
The key idea is to allocate a different class token per task and to average them at the probing stage for downstream tasks.
On four global tasks (music tagging, instrument recognition, key estimation, and pitch estimation), $\text{MT}^2$ matches or outperforms single-token competitors and MERT, a well-known system based on masked language modeling.
While on local tasks, $\text{MT}^2$ trails behind MERT for the beat tracking and ViT-CQT for chord estimation, this may due to the sequence tokens being emergent but not directly optimized. We leave the improvement of these tasks to future work, potentially by further refining the equivariant objectives.
Despite this limitation, our findings demonstrate the value of multitask SSL with multi-class-token transformer in MIR at the 100k-song scale.


\clearpage
\bibliographystyle{IEEEtran}
\bibliography{refs25}

\end{document}